# DENGUE DISEASE PREDICTION USING WEKA DATA MINING TOOL

KASHISH ARA SHAKIL, SHADMA ANIS AND MANSAF ALAM
*Department of Computer Science, Jamia Millia Islamia*
*New Delhi, India*
*shakilkashish@yahoo.co.in, shadamanis@gmail.com and malam2@jmi.ac.in*

Dengue is a life threatening disease prevalent in several developed as well as developing countries like India. This is a virus born disease caused by breeding of Aedes mosquito. Datasets that are available for dengue describe information about the patients suffering with dengue disease and without dengue disease along with their symptoms like: Fever Temperature, WBC, Platelets, Severe Headache, Vomiting, Metallic Taste, Joint Pain, Appetite, Diarrhea, Hematocrit, Hemoglobin, and how many days suffer in different city. In this paper we discuss various algorithm approaches of data mining that have been utilized for dengue disease prediction. Data mining is a well known technique used by health organizations for classification of diseases such as dengue, diabetes and cancer in bioinformatics research. In the proposed approach we have used WEKA with 10 cross validation to evaluate data and compare results. Weka has an extensive collection of different machine learning and data mining algorithms. In this paper we have firstly classified the dengue data set and then compared the different data mining techniques in weka through Explorer, knowledge flow and Experimenter interfaces. Furthermore in order to validate our approach we have used a dengue dataset with 108 instances but weka used 99 rows and 18 attributes to determine the prediction of disease and their accuracy using classifications of different algorithms to find out the best performance. The main objective of this paper is to classify data and assist the users in extracting useful information from data and easily identify a suitable algorithm for accurate predictive model from it. From the findings of this paper it can be concluded that Naïve Bayes and J48 are the best performance algorithms for classified accuracy because they achieved maximum accuracy= 100% with 99 correctly classified instances, maximum ROC = 1 , had least mean absolute error and it took minimum time for building this model through Explorer and Knowledge flow results.

## 1. Introduction

Dengue fever is a disease caused by dengue virus and is also known as break bone fever is transmitted by Aedes mosquito. Infection of dengue is divided into four part DHF I, DHF II, DHF II, DHF IV. It causes life threatening dengue hemorrhagic fever whose symptoms include bleeding, low levels of blood platelets, low blood pressure, metallic taste in mouth, headache, muscle joint pain and rashes.[11, 22]

There is no specific medicine or antibiotic available to treat it. Dengue fever occurs in form of cycles and this cycle is present inside the body of an infected person for two week or less than two week. It causes abdominal pain, hemorrhage (bleeding), and circulatory collapse and Dengue hemorrhagic fever.

The following is the cycle mechanism by which dengue is transmitted: Aedes mosquito carries a virus in its saliva, when it bites a healthy person, that virus enters the person's body and gets mixed with the person's body fluids. The moment white blood vessels gets mixed with the single stranded RNA dengue virus, it starts reproducing inside the white blood cells and thus initiates the dengue virus cycle. In case of severe infection, the duration of virus cycle is prolonged and thereby affects liver and bone marrow leading to less blood circulation in blood vessels, and the blood pressure becomes so low that it cannot supply sufficient blood to all the organs of the body. The bone marrow also does not function properly due to this infection leading to reduced number of platelets and increased risk of bleeding, which are necessary for effective blood clotting.[11]

Bioinformatics research area uses weka tool for solving many data mining problems. Weka stands for Waikato Environment for Knowledge Analysis developed at the university of Waikato in New Zealand and was implemented in 1997 the software freely available at http://www.waikato.ac.nz/ml/weka and written in java language. There are several different levels at which weka can be used. Weka contains modules for data classification and accuracy to predict diseases.[13] Weka has been used in bioinformatics for diagnoses and analysis of dengue disease datasets. Weka has 49 tools for processing, 76 algorithms for classification and regression, 8 algorithms for clustering, and 3 algorithms for finding association rules. Weka algorithms are suitable for generating predictive model accurately by extracting useful information from dengue dataset through WEKA.[4] Apart from weka researchers are now moving towards cloud computing for disease predictions.[23,24,25,28] It also offers facilities such as clustering and analysis of huge datasets.[26,27,29,30] The main focus of this paper is dengue disease prediction using weka data mining tool and its usage for classification in the field of medical bioinformatics. It firstly classifies dataset and then determines which algorithm performs best for diagnosis and prediction of dengue disease. From the findings of the experiments conducted it was revealed that Naïve Bayes and J48 are the best algorithms. The posterior probability of a hypothesis can be estimated using Bayesian reasoning of some given knowledge or data.[31] Prediction begins with identification of symptoms in patients and then identifying sick patients from a lot of sick and healthy ones. Thus, the prime objective of this paper is analysis of data from a dengue dataset classification technique to predict class accurately in each case in data. The major contributions of this paper are:

(1) To extract useful classified accuracy for prediction of dengue diseases.
(2) Comparison of different data mining algorithms on dengue dataset.
(3) Identify the best performance algorithm for prediction of diseases.

In this paper we have used dengue dataset for classification method. The steps followed include collection of the dataset for determining the accuracy, classification and then comparison of results. The dataset has been used to classify the following dengue attributes based on P.I.D, date of dengue fever, days, current temperature, WBC, joint muscles, metallic taste in mouth, appetite, abdomen pain, Nausea, diarrhea and hemoglobin. Several classification algorithms have been used in this paper in order to analyze the performance of applied algorithm on the given dataset but the thrust in this paper is on accuracy measure. Accuracy measures analyze the errors through measures like root mean square error, relative absolute error and correctly classified instances.

Though data mining has several different algorithms to analyze data but analysis using all the methods is not feasible therefore in this paper we have performed the analysis using Naïve Bays, J48 tree, SMO function, REP Tree and Random Tree algorithms by using Explorer, Experimenter and knowledge flow interface of weka tool. The remainder of this paper is organized as follows. Section 2 presents related work done using data mining tools such as weka and CRFSuite to predict diseases; Section 3 describes the methodology adopted by us. Furthermore section 4 presents the Dengue Dataset used in this paper for analysis purpose; Section 5 describes data mining techniques used in this paper for analysis; Result and Discussions are presented in section 6, finally the paper ends with Conclusion and future works in section 7.

## 2. Related Work

Dhamodharan S. has done prediction of liver disease using Bayesian Classification through Naïve Bayes and FT tree algorithms. With the help of data mining techniques they have predicted and analyzed liver diseases using weka tool. They have also compared the outputs obtained from Naïve Bayes and FT tree algorithms and concluded that Naive Bayes algorithm plays a key role in predicting liver diseases.[1]

Solanki A.V. has used weka as a data mining technique for classification of sickle cell disease prevalent in Gujarat. They have compared J48 and Random tree algorithms and have given a predictive model for classification with respect to a person's age of different blood group types. From there experimentation it can be inferred that Random tree is better algorithm as it produces more depth decisions respect to J48 for sickle cell diseases.[2]

Joshi et al. has done diagnosis and prognosis of breast cancer using classification rules. By comparing classification rules such as Bayes Net, Logistic, Multilayer Perceptron, SGD, Simple Logistic, SMO, AdaBoostM1, Attribute Selected, Classification via Regression, Filtered Classifier, Multiclass Classifier and J48,They have inferred that LMT Classifier gives more accurate diagnosis i.e. 76 % healthy and 24 % sick patients.[3]

David S.K. et al. have used classification techniques for leukemia disease prediction. K-Nearest Neighbor, Bayesian Network, Random tree, J48 tree compared on the basis of accuracy, learning time and error rate. According to them Bayesian algorithm has better classification accuracy amongst others.[4]

Vijayarani S. and Sudha S. have compared the analysis of classification function techniques for heart disease prediction. Classification was done using algorithms such as Logistic, Multilayer Perception and Sequential Minimal Optimization algorithms for predicting heart disease. In this classification comparison logistic algorithm trained out to be best classifier for heart disease having more accuracy and least error rate.[5]

Kumar M.N. used alternating decision trees for early diagnosis of dengue fever. The ADTree correctly classifies 84 % of cases as compared to J48 which can classify only 78% of cases correctly.[6]

Durairaj M. and Ranjani V. have compared different data mining applications in healthcare sector. Algorithms such as Naïve, J48, KNN and C4.5 were used for Classification in order to diagnose diseases like Heart Disease, Cancer, AIDS, Brain Cancer, Diabetes, Kidney Dialysis, Dengue, IVF and Hepatitis C. Comparison study analysis revealed high accuracy i.e. 97.77% for cancer prediction and around 70% for IVF treatment through data mining techniques.[7]

Sugandhi C. et al. analyzed a population of cataract patient's database by weka tool. In this study, weka has been used to classify the results and for comparison purpose. They have concluded that Random Tree gives 84% classify accuracy which means better performance as compared to other algorithms used for classification accuracy performance of Naïve Bayes, SMO, J48, REP Tree and Random Tree. Thus according to their study Random Tree is the best performance classification algorithm for cataract patient disease.[8]

Yasodha P. and Kannan M. performed analysis of a population of diabetic patient database using weka tool. They have classified the data and then outputs were compared by using Bayes Network, REP Tree, J48 and Random Tree algorithms. Finally the results conclude that these algorithms help to determine and identify the stage or state in which a of disease like diabetes is in by entering patients daily glucose rate and insulin dosages thereby predicting and consulting the patients for their next insulin dosage .[9]

Bin Othman M.F. and Yau T.M.S. have compared different classification techniques using weka for Breast cancer. In this study they have used different algorithm methods for simulating results of each algorithm and its training. They have simulated the errors by using Bayes Network, Radial Basis function, Decision Tree and pruning and Single Conjugation Rule Learner algorithms. From their work it can be concluded that Bayes Network performs best for breast cancer data. Its time taken to build model is 0.19 second and accuracy 89.7 % and least error at 0.2140 as compared to other algorithms used .[10]

Mihaila C. and Ananiadou S. have compared two data mining tools i.e. weka and CRF Suite on the basis of features like Lexical, Syntactic and semantic with various parameters to compare their impacts on each algorithm. The experiments have been employed in CRF Suite implementation by using Conditional Random Field algorithm and in weka by algorithms like Support Vector machine and Random Forests to identify discourse causality trigger in the biomedical domain. Classification tasks have been performed on the basis of statistics such as F score, precision and recall. As per them CRF is the best performance classifier, achieved F score = 79.35 % by combining three features as compared to other classifier.[21]

Thitiprayoonwonsge D. et al. have analyzed dengue infection using data mining decision tree. In this paper two datasets have been used from two different hospitals Srinagarindra Hospital and Songklanagarind Hospital, each having more than 400 attributes. Four classification algorithms have been used in this paper for experimental purpose. The first and second experiment test got an accuracy of 97.6% and 96.6%. The third experiment extracts useful knowledge. Another objective of this paper was to detect day abatement of fever also referred as day0. In fourth experiment of day0 accuracy is very low as compared to other three experiments. Therefore physician need day0 amongst patient in order to treat them.[22]

## 3. Methodology

In order to carry out experimentations and implementations Weka was used as the data mining tool. Weka (Waikato Environment for Knowledge Analysis) is a data mining tool written in java developed at Waikato. WEKA is a very good data mining tool for the users to classify the accuracy on the basis of datasets by applying different algorithmic approaches and compared in the field of bioinformatics. Explorer, Experimenter and Knowledge flow are the interface available in WEKA that has been used by us. In this paper we have used these data mining techniques to predict the survivability of dengue disease through classification of different algorithms accuracy .[12, 13]

Figure 1 visualizes the interface of WEKA Data mining tool. It has four applications:

(1)Explorer: The explorer interface has several panels like preprocess, classify, cluster, associate, select attribute and visualize. But in this interface our main focus is on the Classification Panel .[12]

(2) Experimenter: This interface provides facility for systematic comparison of different algorithms on basis of given datasets. Each algorithm runs 10 times and then the accuracy reported.[12]

(3)Knowledge Flow: It is an alternative to the explorer interface. The only difference between this and others is that here user selects Weka component from toolbar and connects them to make a layout for running the algorithms .[12]

(4) Simple CLI: Simple CLI means command line interface. User performs operations through a command line interface by giving instructions to the operating system. This interface is less popular as compared to other three.

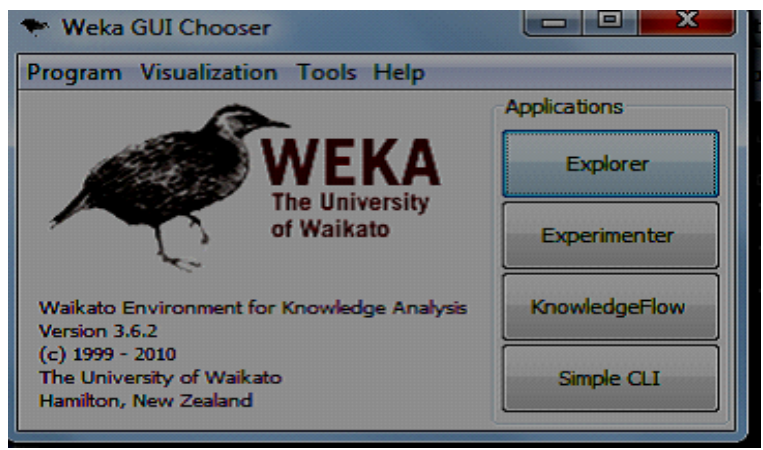

Fig. 1. Screenshot view of WEKA

## 3.1 Classification

In data mining tools classification deals with identifying the problem by observing characteristics of diseases amongst patients and diagnose or predict which algorithm shows best performance on the basis of WEKA's statistical output .[20]

Table 1 shows the WEKA data mining techniques that have been used in this paper along with other prerequisites like data set format etc. by using different algorithms.

Table 1.Weka data mining technique by using different algorithms

| Software | Datasets | Weka Data Mining Technique | Classification Algorithms | Operating System | Dataset File Format | Purpose |
|---|---|---|---|---|---|---|
| WEKA | Dengue | Explorer Experimenter | Naïve Bayes J48 SMO REP Random | Windows 7 | CSV | Classification |

Three techniques have been adopted in this paper, the first technique uses explorer interface and depends on algorithms like Naïve Bayes, SMO, J48, REP Tree and RANDOM Tree, used in areas to represent, utilize and learn the statistical knowledge and significant results have been achieved.

The second technique uses Experimenter interface. This study allows one to design experiments for running algorithms such as Naïve Bayes, J48, REP Tree and RANDOM Tree on datasets. These algorithms can be run on experimenter and analyze the results. It configures the test option to use cross validation 10 folds. This interface provides provision for running all the algorithms together and thus a comparative result was obtained.

The third technique uses Knowledge Flow. In this study we classified the accuracy of different algorithms Naïve Bayes, SMO, J48, REP Tree and random Tree on different data sets and compared the results to know which algorithm shows best performance. In order to predict Dengue Disease for survivability by user one can select this weka component from toolbar, place them in a layout like manner and connect its different components together in order to form a knowledge flow web for preprocessing and analyzing data.

All the algorithms used by us were applied to a dengue data set explained in detail in section 4. In order to obtain better accuracy 10 fold cross validation was performed. For each classification we selected training and testing sample randomly from the base set to train the model and then test it in order to estimate the classification and accuracy measure for each classifier. The thrust classifications and accuracy used by us are:

*3.1.1. Correctly Classified Accuracy*

It shows the accuracy percentage of test that is correctly classified.

*3.1.2. Incorrectly Classified Accuracy*

It shows the accuracy percentage of test that is incorrectly classified.

*3.1.3. Mean Absolute Error*

It shows the number of errors to analyze algorithm classification accuracy.

*3.1.4. Time*

It shows how much time is required to build model in order to predict disease.

*3.1.5. ROC Area*

Receiver Operating Characteristic19 represent test performance guide for classifications accuracy of diagnostic test based on: excellent (0.90-1), good (0.80-0.90), fair (0.60-0.70), poor (0.60-0.70), fail (0.50 – 0.60).

**4. Datasets Used**

Dataset is a collection of data or a single statistical data where every attribute of data represents variable and each instance has its own description. For prediction of dengue disease we used dengue data set for prediction and classification of algorithms in order to compare their accuracy using wekas three interfaces: Explorer, Experimenter and Knowledge Flow.

Figure 2 shows a description of dengue dataset. The dataset used by us contains 18 attributes and 108 instances for dengue disease classification and accuracy. We have applied different algorithms using WEKA data mining tool for our analysis purpose.14

Fig. 2. Screenshot view of Dengue Dataset

Table 2. Describes the attributes of data set which are presented in Figure 2 .The file format of datasets used is Comma Separated Value CSV. Each attribute shows the present absent of dengue symptoms, number of days , date , number of WBC, number of platelets, pain and taste among patients in different cities and how many days they suffers .

Table 2 . Description of datasets attributes

| Attributes | Description |
| --- | --- |
| P.I.D | Patient ID |
| Date of fever | Month |
| Residence | City |
| Days | No. of days |
| Current Temperature | Fever |
| WBC | No. of WBC |
| Severe Headache | Yes or No |
| Pain | Behind Eyes |
| Joint / Muscle pain | Yes or No |
| Metallic Taste | Yes or No |
| Appetite | Yes or No |
| Abdominal pain | Yes or No |
| Nausea/Vomiting | Yes or No |
| Diarrhea | Yes or No |
| Hemoglobin | Hemoglobin Range |
| Hematocrit | Hematocrit Range |
| Platelets | No. of Platelets |
| Dengue | Yes or No |

## 5. Data Mining Techniques

The data mining technique have been used by us to predict dengue disease. Predictions have been done by us using weka data mining tool for classification and accuracy by applying different algorithms approaches. The interfaces of weka used in this paper are the following:

### *5.1. Explorer Interface*

It first preprocesses the data and then filters the data. Users can then load the data file in CSV (Comma Separated Value) format and then analyze the classification accuracy result by selecting the following algorithms using 10 cross validation: Naïve Bayes, J48, SMO, REP Tree, and Random Tree. Figure 3 shows the interface of explorer when using dengue dataset is opened using CSV file along with its graphical view.[12, 15]

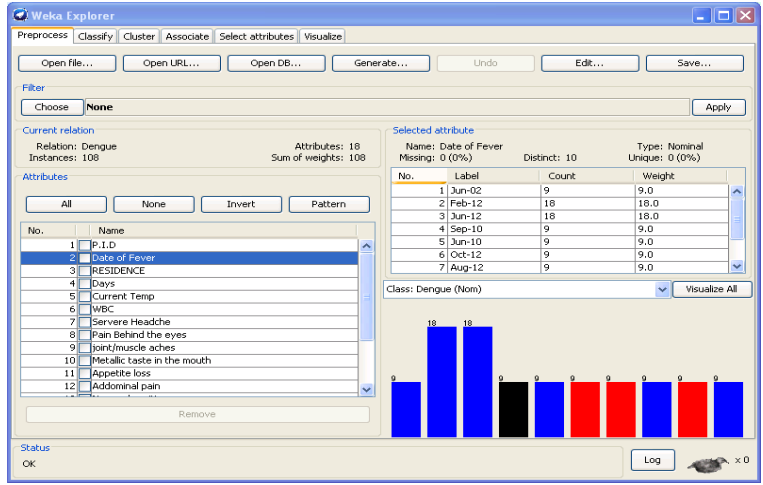

Fig. 3. Screenshot view of CSV Dengue Dataset File open in Explorer Interface

### 5.1.1. Naïve Bayes

Naïve Bayes is one of the algorithms that works as a probabilistic classifier of all attributes contained in data sample individually and then classifies data problems. Running the algorithms using Naïve Bayes we analyze the classifier output with so many statistics based output by using 10 cross validation to make a prediction of each instance of the dataset.[16]

After running these algorithms we achieved a classification accuracy of 100% for 99 correctly classified instances, error rates achieved i.e. Mean Absolute Error is 0.0011, time taken for building model is 0 seconds and ROC area is 1these outputs are obtained after these algorithms are run .

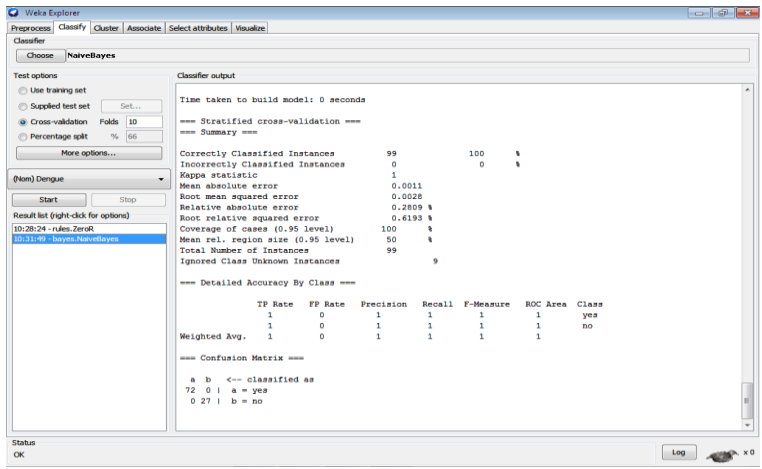

Fig. 4. Screenshot view for Naïve Bayes Algorithm

The output obtained by scoring of NaïveBayes algorithm accuracy of is given in Table 3 on the basis of time, accuracy, error and ROC.

Table 3: Naïve Bayes algorithm accuracy

| Algorithm | Time Taken to Build Model (seconds) | Correctly Classified Instances %Accuracy | Incorrectly Classified Instances %Accuracy | Mean Absolute Error | ROC Area |
|---|---|---|---|---|---|
| Naïve Bayes | 0 | 100% (99) | 0% (0) | 0.0011 | 1 |

### 5.1.2. J48 Tree

J48 Tree has been used in this paper to decide the target value based on various attributes of dataset to predict machine learning model and classify their accuracy. We have also used J48 Tree on our dengue disease dataset. After running this algorithm we analyzed the outputs obtained from the classifier, the output gave several statistics based on 10 cross validation to make a prediction of each instances of dataset. Figure 5 shows the classification accuracy achieved from this algorithm i.e. 100% is the correctly classified accuracy for a batch of 99 instances, mean absolute error obtained is 0, time taken to build this model is 0 seconds, and ROC area is 0.958.

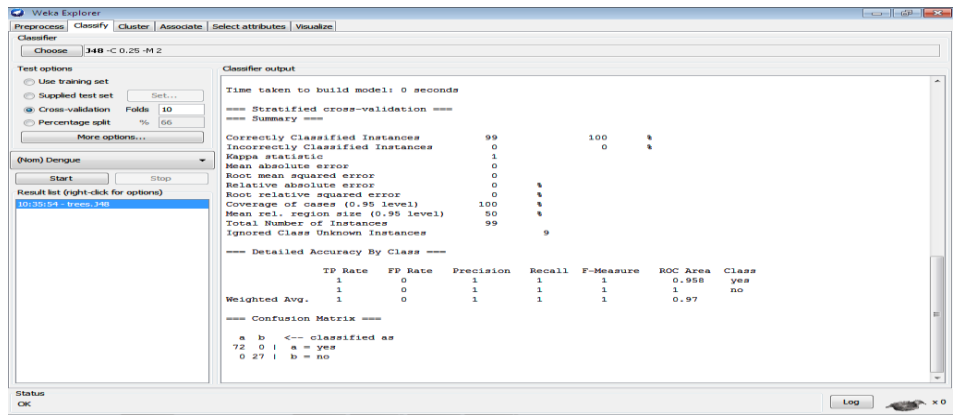

Fig. 5. Screenshot view J48 Algorithm

Scoring by using J48 algorithm accuracy is given in Table 4 on the basis of of time, accuracy, error and ROC.

Table 4. J48 algorithm accuracy

| Algorithm | Time Taken to Build Model (seconds) | Correctly Classified Instances %Accuracy | Incorrectly Classified Instances %Accuracy | Mean Absolute Error | ROC Area |
|---|---|---|---|---|---|
| J48 | 0 | 100% (99) | 0% (0) | 0 | 0.958 |

*5.1.3. SMO*

SMO is one of the methods used for classification. In this paper we have used this algorithm to split the data on the basis of dataset. Running this algorithm we analyzed the classifier output with different statistics based on output by using 10 cross validation to make a prediction of each instances of dataset.

Figure 6 shows the classification accuracy of 100%, error rates that is mean absolute error obtained is 0, time taken to build model is 0 seconds and ROC area is 0.875 that is obtained after running these algorithms.

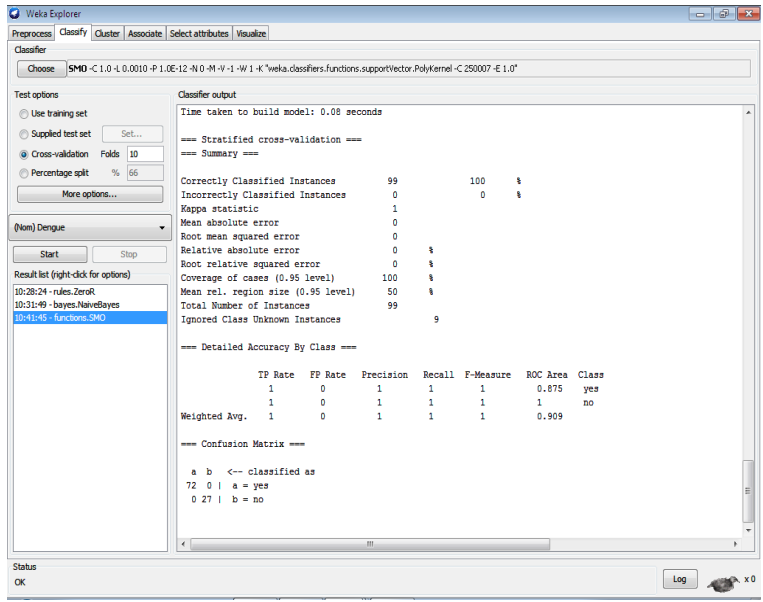

Fig. 6. Screenshot view SMO Algorithm

Scoring of J48 Algorithm Accuracy is given in Table 5 on the basis of of time, accuracy, error and ROC.

Table 5. SMO algorithm accuracy

| Algorithm | Time Taken to Build Model (seconds) | Correctly Classified Instances %Accuracy | Incorrectly Classified Instances %Accuracy | Mean Absolute Error | ROC Area |
|---|---|---|---|---|---|
| SMO | 0 | 100% (99) | 0% (0) | 0 | 0.875 |

*5.1.4. REP Tree*

REP Tree has been used in this paper to build a decision and reduces errors by sorted values of numeric attribute and splits the instances into pieces to classify the accuracy. Running the algorithm we analyze the classifier output with statistics based outputs by using 10 cross validation to make a prediction of each instance of dataset.

In figure 7 classification accuracy achieved shows that 74.7475 % are correctly classified accuracy for 74 instances, 25.2525 % incorrectly classified accuracy for 25 instances, error rates that is mean absolute error is 0.3655, time taken to build model is 0.02 seconds and ROC area is 0.544 these are mentioned in output.

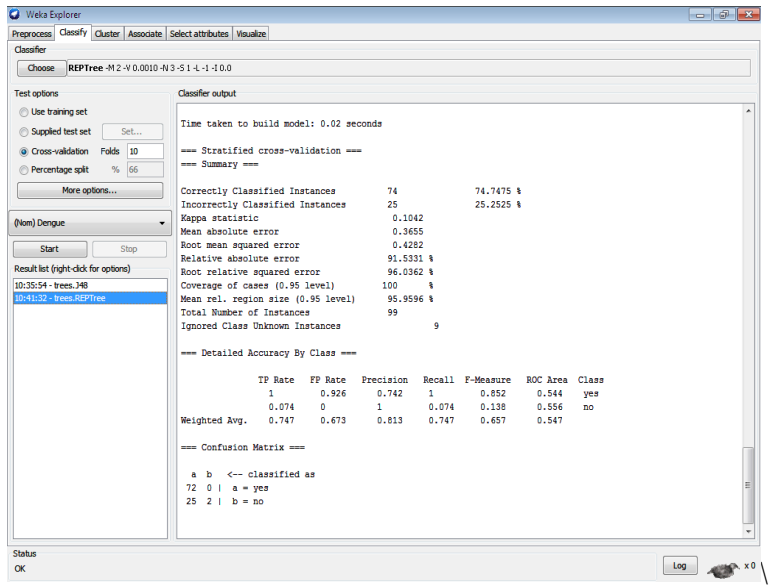

Fig. 7. Screenshot view of REP Tree Algorithm

Figure 7 analyzes scoring of REP Tree Algorithm Accuracy and the output is elaborated further in Table 6 on the basis of time, accuracy, error and ROC.

Table 6 : REP Tree algorithm accuracy

| Algorithm | Time Taken to Build Model (seconds) | Correctly Classified Instances %Accuracy | Incorrectly Classified Instances %Accuracy | Mean Absolute Error | ROC Area |
|---|---|---|---|---|---|
| REP Tree | 0.02 | 74.7475% (74) | 25.2525% (25) | 0.3655 | 0.544 |

### 5.1.5. Random Tree

Random Tree has been used in this paper for randomly choosing k attributes at each node to allow the estimation of class probabilities. Running the algorithm we analyze the classifier output with statistics based output by using 10 cross validation to make a prediction of each instances of dataset.

From figure 8 classification accuracy of 87.8788% is obtained, it is correctly classified accuracy for 87 instances, 12.1212% incorrectly classified accuracy for 12 instances, error rates that is mean absolute error is 0.1853, time taken to build this model is 0 seconds, and ROC area is 0.876 these are mentioned in output.

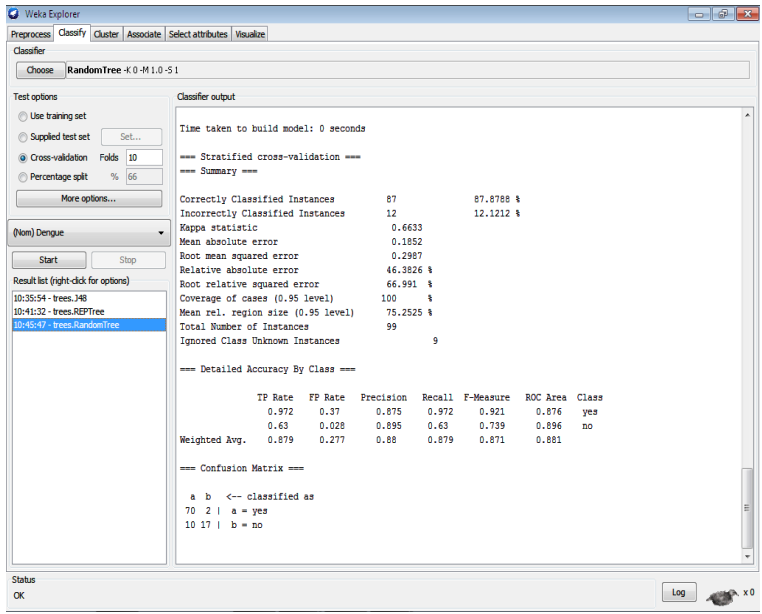

Fig. 8. Screenshot view of Random Tree Algorithm

Scoring output obtained by using random tree algorithm accuracy is given in Table 7 on the basis of time, accuracy, error and ROC.

Table 7. Random Tree algorithm accuracy

| Algorithm | Time Taken to Build Model (seconds) | Correctly Classified Instances %Accuracy | Incorrectly Classified Accuracy | Mean Absolute Error | ROC Area |
|---|---|---|---|---|---|
| Random Tree | 0 | 87.8788% (87) | 12.1212% (12) | 0.1853 | 0.876 |

### 5.2. *Experimenter Interface*

Experimenter Interface has been used in this paper to analyze data by experimenting through algorithms such as Naïve Bayes, J48, REP Tree and Random Tree to classify the data using train and test sets.[17]

In Figure 9 we run four different algorithms on dengue datasets and analyze algorithms accuracy.

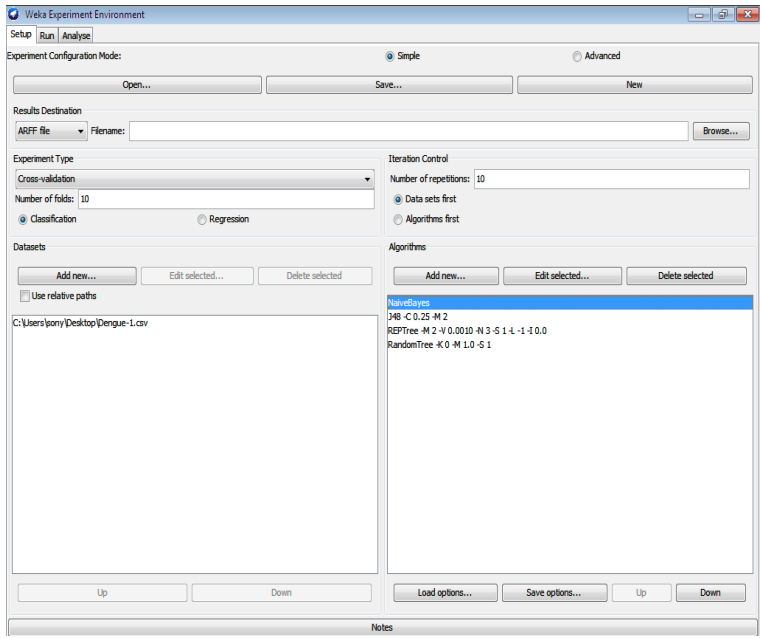

Fig. 9. Screenshot view of Experimenter Interface

*5.2.1. Naïve Bayes*

It is one of the fastest algorithm works on probability of all attribute contained in data sample individually and then classifies them accurately.

*5.2.2. J48 Tree*

We used J48 tree to decide the target value based on various attribute of dataset to predict algorithms accuracy.

*5.2.3. REP Tree*

We used Weka classifier tree algorithm analyze accuracy applied on dengue dataset.

*5.2.4. Random Tree*

We used Random classifier tree algorithm to analyze classification based on our dataset. Figure 10 analyzes experiment test of all four algorithms, each algorithm is run 10 times and accuracy is reported. "v" stand for best accuracy prediction and "*" stand for worse accuracy prediction. This means it predicts best and worse scoring accuracy amongst the four different algorithms listed below respectively:
- Naïve Bayes
- J48 Tree
- REP Tree
- Random Tree

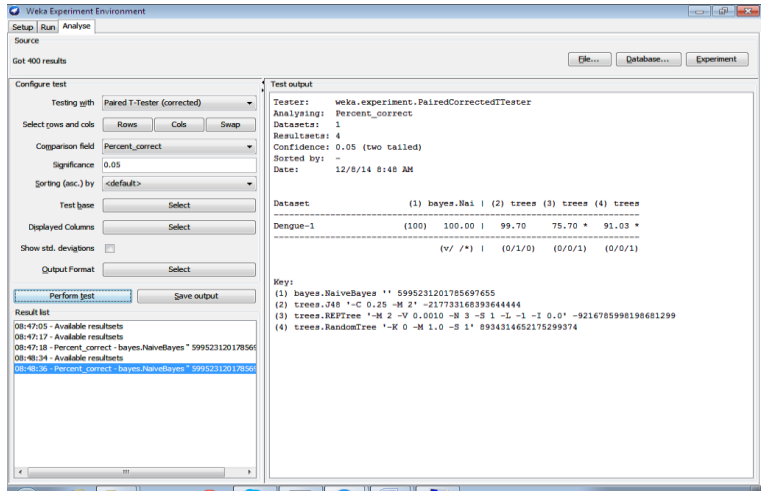

Fig. 10.  Screenshot view of Experimenter Algorithm Accuracy

Scoring accuracy of Naïve Bayes, J48, REP Tree and Random Tree is given in Table 8

Table 8. Experimenter algorithms accuracy

| Algorithm | Best Accuracy Prediction (v) | Worse Accuracy Prediction (*) |
|---|---|---|
| Naïve Bayes | 100% | _ |
| J48 Tree | 99.70% | _ |
| REP Tree | _ | 75.70% |
| Random Tree | _ | 91.03% |

### *5.3. Knowledge Flow Interface*

Knowledge Flow is an alternative to the explorer.[18] the user lays out the data by connecting them together in order to form a knowledge flow by selecting weka component from a tool bar as shown in Figure 11. For the purpose of our experimentation we have connected together CSV loader, class assigner, Cross validation, and then an algorithm such as SMO, REP tree etc followed by Classifier Performance evaluator and finally we view the output using text viewer.

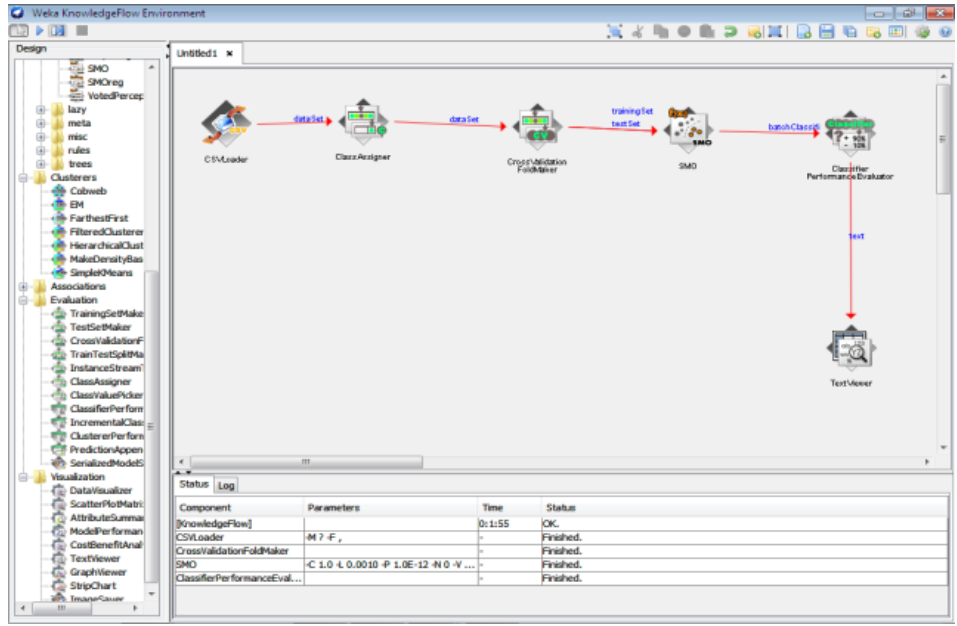

Fig. 11. Screenshot view of Knowledge Flow Interface

*5.3.1. Naïve Bayes*

Naïve Bayes is the algorithm works on probability of all attribute contained in data sample individually and classify data problems. We used this algorithm in Experimenter interface also.

In figure 12classification accuracy achieved is 100% correctly classified accuracy for 99 instances, error rate that is mean absolute error is 0.0011, time taken to build model is 0 seconds and ROC area is 1 these are mentioned in output.

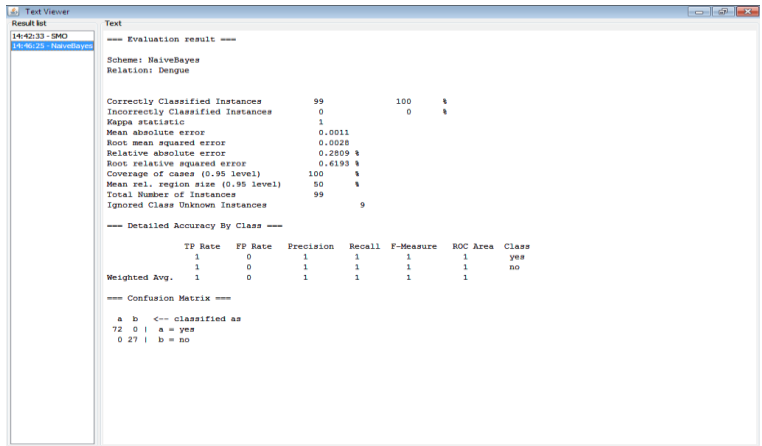

Fig. 12.  Screenshot view of Naïve Bayes Algorithm

Figure 12 analyses scoring of Naïve Bayes algorithm accuracy given in Table 9 on the basis of time, accuracy, error and ROC are the following:

Table9. Naïve Bayes algorithm accuracy

| Algorithm | Time Taken to Build Model (seconds) | Correctly Classified Instances %Accuracy | Incorrectly Classified Instances %Accuracy | Mean Absolute Error | ROC Area |
|---|---|---|---|---|---|
| Naïve Bayes | 0 | 100% (99) | 0% (0) | 0.0011 | 1 |

*5.3.2. J48 Tree*

J48 Tree decides the target value based on various attributes of dataset to predict machine learning model and classify their accuracy on the basis of dengue disease dataset.

In figure 13 classification accuracy achieved shows that 100% are correctly classified accuracy out of 99 instances , error rates that is mean absolute error is 0, time taken to build model is 0 seconds and ROC area is 1 these are mentioned in output.

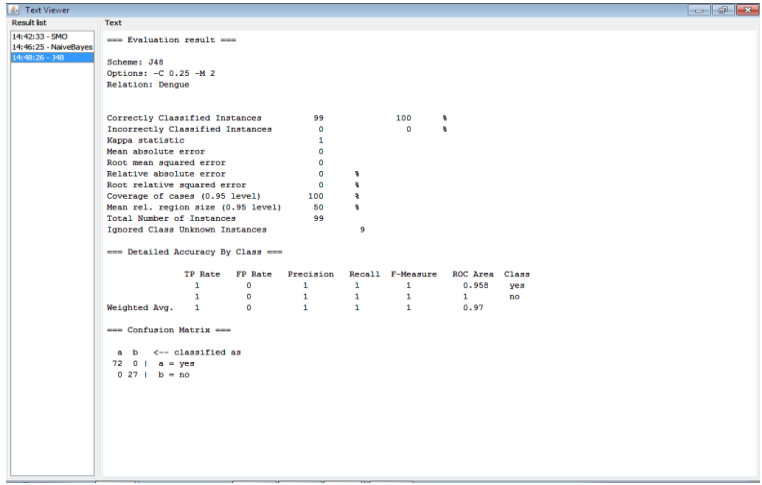

Fig. 13.  Screenshot view of J48 Tree Algorithm

A detailed analysis of scoring using J48 Algorithm is given by Table 10 on the basis of of time, accuracy, error and ROC.

Table 10.J48 Tree algorithm accuracy

| Algorithm | Time Taken to Build Model (seconds) | Correctly Classified Instances %Accuracy | Incorrectly Classified Instances %Accuracy | Mean Absolute Error | ROC Area |
|---|---|---|---|---|---|
| J48 | 0 | 100% (99) | 0% (0) | 0 | 1 |

*5.3.3. SMO*

SMO algorithm has also been used by us in knowledge flow interface for classification. It splits the data on the basis of dataset and then analyses the output. From figure 14  we can deduce that classification accuracy achieved gives 100% correctly classified accuracy out of 99 instances, error rates that is mean absolute error is 0, time taken to build model is 0 seconds, and ROC area is 0.875.

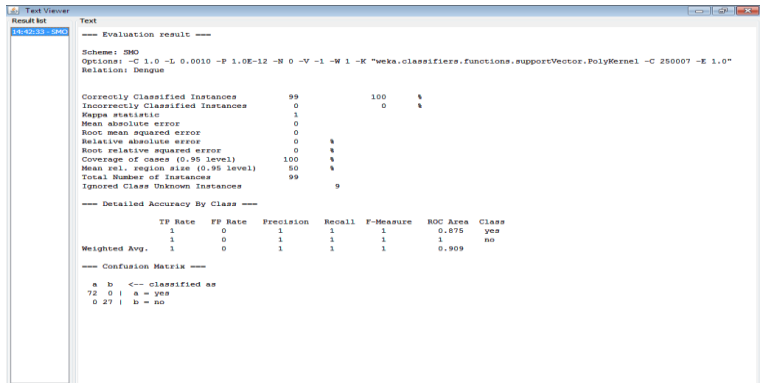

Fig. 14. Screenshot view of SMO Algorithm

Scoring of SMO algorithm accuracy is shown by Table 11 on the basis of time, accuracy, error and ROC.

Table 11 .SMO algorithm accuracy

| Algorithm | Time Taken to Build Model (seconds) | Correctly Classified Instances %Accuracy | Incorrectly Classified Instances %Accuracy | Mean Absolute Error | ROC Area |
|---|---|---|---|---|---|
| SMO | 0 | 100% (99) | 0% (0) | 0 | 0.875 |

### 5.3.4. REP Tree

REP Tree has been used in this paper to build a decision tree and thereby reduce errors by sorting values of numeric attributes and splits instances into pieces to classify the accuracy.

Figure 15shows that classification accuracy achieved is 74.7475% correctly classified accuracy , 25.2525% are incorrectly classified accuracy, error rates that is mean absolute error is 0.3655,time taken to build model is 0.02 and ROC area is 0.544.

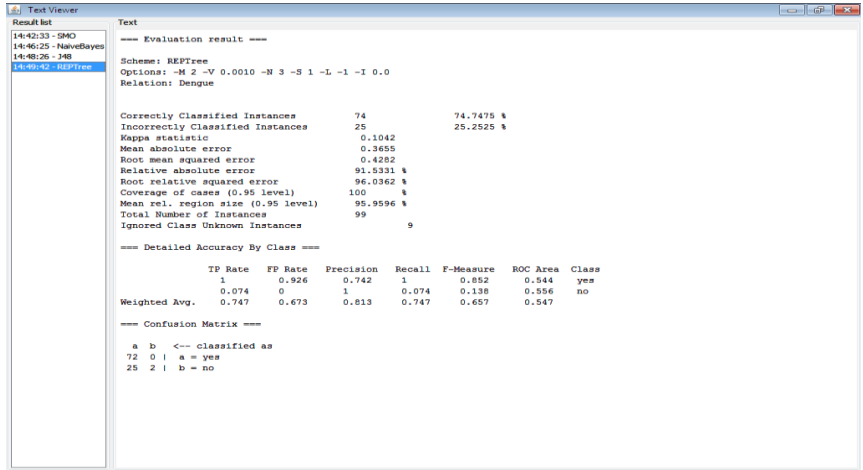

Fig. 15. Screenshot view of REP Tree Algorithm

Table 12 shows the analysis of scoring of REP tree algorithm accuracy on the basis of time, accuracy, error and ROC.

Table 12. REP Tree algorithm accuracy

| Algorithm | Time Taken to Build Model (seconds) | Correctly Classified Instances %Accuracy | Incorrectly Classified Instances %Accuracy | Mean Absolute Error | ROC Area |
|---|---|---|---|---|---|
| REP Tree | 0.02 | 74.7475% (74) | 25.2525% (25) | 0.3655 | 0.544 |

*5.3.5. Random Tree*

Random Tree randomly chooses attributes at each node to allow the estimation of class accuracy

From figure 16 we can observe that classification accuracy achieved is 87.8788% correctly classified accuracy, 12.1212% are incorrectly classified accuracy, error rates that is mean absolute error is 0.1853, time taken to build model is 0 seconds and ROC area is 0.876 these are mentioned in output.

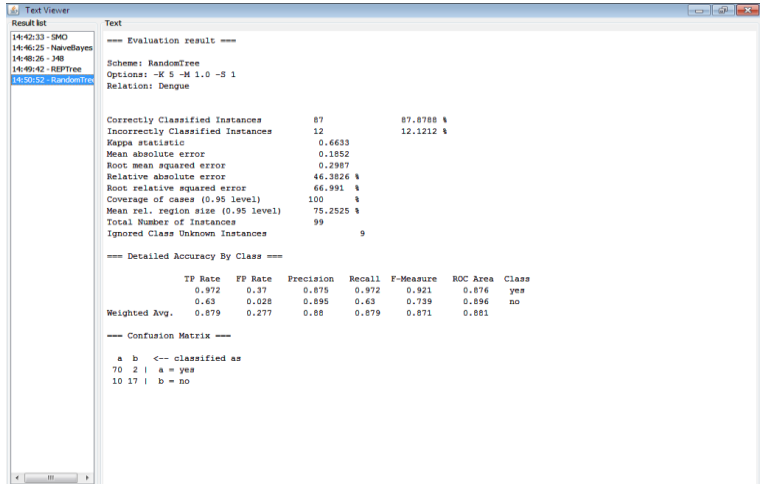

Fig. 16. Screenshot view of Random Tree Algorithm

Figure 16 analyses scoring of J48 algorithm accuracy they are illustrated further in Table 13 on the basis of time, accuracy, error and ROC.

Table 13 : Random Tree algorithm accuracy

| Algorithm | Time Taken to Build Model (seconds) | Correctly Classified Instances %Accuracy | Incorrectly Classified Instances %Accuracy | Mean Absolute Error | ROC Area |
|---|---|---|---|---|---|
| Random Tree | 0 | 87.8788% (87) | 12.1212% (12) | 0.1853 | 0.876 |

## 6. Results and Discussion

Explorer, Experimenter and Knowledge flow are the data mining techniques that have been used by us using different algorithms Naïve Bayes, J48, SMO, RANDOM tree and REP tree. Through these techniques we trained out results on the basis of time taken to build model, correctly classified instances, error and ROC area. Algorithm scoring accuracy is shown in Table 14. Naïve Bayes and J48 classified 100 % correctly instances accuracy with minimum Naïve Bayes Mean Absolute Error = 0.011 and Naïve Bayes Mean Absolute Error J48=0, having maximum Naïve Bayes ROC =1 and J48ROC Area = 0.958 and time taken to build model=0 seconds. So from Explorer Interface data mining technique we can deduce that Naïve Bayes and J48 have maximum accuracy , least error and it takes less time to build model it and has maximum ROC.

Table 14. Explorer result

| Algorithm | Time Taken to Build Model (seconds) | Correctly Classified Instances %Accuracy | Incorrectly Classified Instances %Accuracy | Mean Absolute Error | ROC Area |
|---|---|---|---|---|---|
| NaïveBayes | 0 | 100%(99) | 0%(0) | 0.0011 | 1 |
| J48 | 0 | 100%(99) | 0%(0) | 0 | 0.958 |
| SMO | 0 | 100%(99) | 0%(0) | 0 | 0.875 |
| REP Tree | 0.02 | 74.74%(74) | 25.25%(25) | 0.3655 | 0.544 |
| Random Tree | 0 | 87.87%(87) | 12.12%(12) | 0.1853 | 0.876 |

In Table 15 Naïve Bayes and J48 classified 100 % correctly instances accuracy with minimum Naïve Bayes Mean Absolute Error = 0.011 and Naïve Bayes Mean Absolute Error J48=0, having maximum Naïve Bayes ROC =1 and J48 ROC Area = 0.958 and time taken to build model=0 seconds. So from Knowledge flow Interface data mining technique result Naïve Bayes and J48 have maximum accuracy, least error, less time taken to build model and maximum ROC. Explorer and Knowledge flow achieved same scoring to classify accuracy but there is approx. change in ROC Value of Naïve Bayes and J48 as compared to other because Knowledge flow is an alternative method of Explorer.

Table 15. Knowledge Flow result

| Algorithm | Time Taken to Build Model (seconds) | Correctly Classified Instances %Accuracy | Incorrectly Classified Instances %Accuracy | Mean Absolute Error | ROC Area |
|---|---|---|---|---|---|
| Naïve Bayes | 0 | 100% (99) | 0% (0) | 0.0011 | 1 |
| J48 | 0 | 100%(99) | 0% | 0 | 1 |
| SMO | 0 | 100%(99) | 0%(0) | 0 | 0.875 |
| REP Tree | 0.02 | 74.74%(74) | 25.25%(25) | 0.3655 | 0.544 |
| Random Tree | 0 | 87.87%(87) | 12.12%(12) | 0.1853 | 0.876 |

In Table 16 Naïve Bayes and J48 scoring accuracy is high that is best prediction (V) as compared to REP Tree and Random Tree having low algorithm accuracy called worse prediction (*).

Table 16 : Experimenter result

| Algorithm | Best Accuracy Prediction (v) | Worse Accuracy Prediction (*) |
|---|---|---|
| Naïve Bayes | 100% | _ |
| J48 Tree | 99.70% | _ |
| REP Tree | _ | 75.70% |
| Random Tree | _ | 91.03% |

Finally from these three data mining technique it is observed that Naïve Bayes and J48 are the best classifier performance to predict the survivability of dengue disease prediction among patient using WEKA because it classifies more accurately, has maximum ROC Area, least mean absolute error and takes minimum time to build model . The Accuracy of test depends on dataset with and without disease. Accuracy measured by ROC area = 1 shows a perfect and excellent test as Patient will get effective diagnosis timely and in an accurate manner.

**7. Conclusion and Future Work**

The main aim of this paper is to predict dengue disease using WEKA data mining tool. It has four interfaces. Out of these four we have used three interfaces: Explorer, Experimenter and knowledge flow. Each interface has its own classifier algorithms. We have used five algorithms i.e. Naïve Bayes, J48, SMO, REP Tree and Random tree for our experimentation. Then these algorithms were implemented using WEKA data mining technique to analyze algorithm accuracy which was obtained after running these algorithms in the output window. After running these algorithms the outputs were compared on the basis of accuracy achieved. In Explorer and Knowledge flow there are several scoring algorithms for accuracy but for our experimentation we have used only five algorithms. The outputs obtained from both Explorer and Knowledge flow is approximately same because knowledge flow is an alternative method of Explorer. It is just a different way of carrying out experimentations. These algorithms compare classifier accuracy to each other on the basis of correctly classified instances, time taken to build model, mean absolute error and ROC Area.

Through Explorer and Knowledge Flow technique it was inferred that Nave Bayes and J48 are the best performance classifier algorithms as they achieved an accuracy of 100 %, takes less time taken to build and shows maximum ROC area = 1, and had least absolute error. Maximum ROC Area means excellent predictions performance as compared to other algorithms.

Experimenter result showed that scoring accuracy of Naïve Bayes is 100% and J48 is 99.70% as compared to REP tree and Random tree so we can conclude that in Experimenter interface Naïve Bayes and J48 are the best classifier algorithms for accuracy predictions for dengue disease survivability on the basis of symptoms given in dataset among patients.

The applications of Weka can be extended further to medical field for diagnosis of different diseases like cancer and many others. It can also help in solving the problems of clinical research using different applications of Weka. Another advantage of using Weka for prediction of diseases is that it can easily diagnose a disease even in case when the number of patients for whom the prediction has to be done is huge or in case of very large data sets spanning lakhs of patients. Even though Weka is a powerful data mining tool to analyze the overview of classification, clustering, Association Rule Mining and visualization of result in medical health to predict disease among patient but we can use other tools such as Matlab in order to further classify different data sets .The proposed approach is used with dengue data set but we plan to extend this approach in future for prediction of other diseases such as cancer etc.

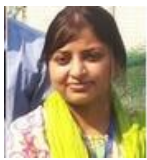
**Kashish Ara Shakil** has received her Bachelor's degree in Computer Science from Delhi University and also holds an MCA degree as well. She is currently pursuing her doctoral studies in Computer Science from Jamia Millia Islamia (A Central University). She has written several research papers in the field of Cloud computing. Her area of interest includes database management system, cloud computing and parallel and distributed computing

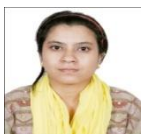
**Shadma Anis** has received her bachelor's degree in Life Sciences form Delhi University in 2012 and then did her Masters in Bioinformatics from Jamia Millia Islamia (A Central University); India in 2012 -2014 .Her area of interest includes Data mining, cloud computing, proteomics and drug design.

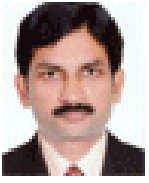

**Mansaf Alam** received his doctoral degree in computer Science from Jamia Millia Islamia New Delhi in the year 2009.  He is currently working as an Assistant. Professor in the Department of Computer Science, Jamia Millia Islamia. He is also the Editor-in-Chief, Journal of Applied Information Science. He is in editorial Board of some reputed International Journals in Computer Sciences and has published about 24 research papers. He also has a book entitled as "Concepts of Multimedia, Book" to his credit. His areas of research include Cloud database management system (CDBMS), Object Oriented Database System (OODBMS), Genetic Programming, Bioinformatics, Image Processing, Information Retrieval and Data Mining.

.